\documentclass[ ]{copernicus}

\usepackage{url,lineno}
\linenumbers
\usepackage{color}
\usepackage[T1]{fontenc}

\begin{document}

\title{Superdiffusion revisited {in view of collisionless reconnection}
}

\author[1,2]{R. A. Treumann\thanks{Visiting the International Space Science Institute, Bern, Switzerland}}
\author[3]{W. Baumjohann}

\affil[1]{Department of Geophysics and Environmental Sciences, Munich University, Munich, Germany}
\affil[2]{Department of Physics and Astronomy, Dartmouth College, Hanover NH 03755, USA}
\affil[3]{Space Research Institute, Austrian Academy of Sciences, Graz, Austria}

\runningtitle{Superdiffusion}

\runningauthor{R. A. Treumann and W. Baumjohann}

\correspondence{R. A.Treumann\\ (rudolf.treumann@geophysik.uni-muenchen.de)}

\received{ }
\revised{ }
\accepted{ }
\published{ }


\firstpage{1}

\maketitle

\begin{abstract}
The concept of diffusion in collisionless space plasmas like those near the magnetopause and in the geomagnetic tail during reconnection is reexamined making use of the division of particle orbits into waiting orbits and break-outs into ballistic motion lying at the bottom, for instance, of L\'evy flights. The rms average displacement in this case increases with time, describing superdiffusion, though faster than classical, is still a weak process, being however strong enough for supporting fast reconnection. Referring to two kinds of numerical particle-in-cell simulations we determine the anomalous diffusion coefficient, the anomalous collision frequency on which the diffusion process is based, and construct a relation between the diffusion coefficients and the resistive scale. The anomalous collision frequency from electron pseudo-viscosity in reconnection turns out of being of the order of the lower-hybrid frequency with the latter providing a lower limit, thus making similar assumptions physically meaningful. Tentative though not completely justified use of the $\kappa$ distribution yields $\kappa\approx 6$ in the reconnection diffusion region, and the anomalous diffusion coefficient the order of several times Bohm diffusivity.

\end{abstract}

\introduction
Anomalous diffusion is the summary heading of all processes where the ensemble averaged mean-square displacement $\langle x^2\rangle\propto t^\gamma$ deviates from linear time dependence $\gamma=1$ with classical (Einstein) diffusion coefficient $D_\mathit{cl}= 2T \nu_c/m$, with $T$ temperature, and $\nu_c$ classical binary collision frequency. For $\gamma>1$ one speaks of superdiffusion, which is of particular importance in the collisionless space plasma where classical diffusion is practically inhibited on all physically interesting processes. (The less interesting case $\gamma<1$ would refer to subdiffusion.) One of those processes is reconnection,  the dominant mechanism for plasma and magnetic field transport across magnetic boundaries represented by thin current sheets/layers. 

Reconnection has the enormous advantage over global diffusion of being localized, with the main physics of magnetic merging and plasma mixing taking place in an extraordinarily small spatial region the linear size shorter than the electron inertial length $<\lambda_e=c/\omega_e$. In this note, based on available numerical simulations, we demonstrate by estimating the anomalous collision frequency $\nu_a$ that magnetic merging during reconnection can well be understood as a localized anomalous diffusion process. This result satisfactorily unifies the two originally different views on plasma transport across an apparently impermeable boundary like the magnetopause.

Anomalous diffusion is also of interest in cosmic ray physics, where it is frequently described as quasilinear diffusion resulting from wave-particle interactions, formulated in the Fokker-Planck phase space-diffusion formalism. Unfortunately, most of the observed diffusive particle spectra \citep[cf., e.g.][for the most elaborate observations in near-Earth space]{christon1989,christon1991} barely exhibit the shapes resulting from quasilinear diffusion. They turn out power law both in energy and momentum space, most frequently being described best by so-called $\kappa$-distributions
\begin{equation}\label{eq-two}
p(\kappa\,|\,\mathbf{x})=A_\kappa\left(1+\frac{\mathbf{x}^2}{\kappa \ell^2}\right)^{-(\kappa +1+d/2)}
\end{equation}
with normalization factor $A$, $d$ dimensionality, and $\ell$ correlation length \citep[cf., e.g.,][for an almost complete compilation of the properties of $\kappa$-distributions]{liva2010,liva2011,liva2013}  with high-energy/high-momentum slopes to which the parameters $\kappa$ are related. Estimated $\kappa$ values from the magnetospheric observations range in the interval $5<\kappa<10$ \citep{christon1991}. Such distributions were introduced by \cite{vasyliunas1968}, following a suggestion by S. Olbert, as best fits.\footnote{Theoretical attempts of justifying solar wind $\kappa$-distributions followed, invoking wave-particle interactions with inclusion of residual binary collisions \citep{scudder1979}. Statistical mechanical arguments were based on nonextensive statistical mechanics \citep{tsallis1988,gell-mann2004}. From kinetic theory they were identified as collisionless turbulent quasi-stationary states far from thermal equilibrium resulting from anomalous wave-particle interactions \citep{treumann1999a,treumann1999b}. There  the role of the temperature $T$ as thermodynamic derivative was clarified \citep[see also][]{liva2010}. The relation between the nonextensive $q$ and the $\kappa$ parameters was given first in \citet{treumann1997}.} In the time-asymptotic limit, $\kappa$ distributions were explicitly derived by \citet{hasegawa1985} and \citet{yoon2012}. Their $q$-equivalent relation to superdiffusion has also been suggested \citep[][and references therein]{tsallis1995,prato1999,bologna2000,gell-mann2004}. 

For the present puroposes we make no direct use of these distributions as they, apparently, play no role in reconnection. Rather, as we demonstrate, anomalous diffusion in reconnection results from processes leading to waiting statistics and causing gyro-viscosity.

\section{Diffusion process}

Collisionless dissipation and related diffusion is mediated in a wider sense by collisionless turbulence \citep[cf., e.g.,][]{allegrini1996}. Here binary collision times $\tau_c\gg\tau_a$ by far exceed anomalous interaction times. Any real non-collisional diffusion proceeds on times much shorter than classical (in comparison infinite) diffusion times with absolute values of anomalous diffusion coefficients being small.

The superdiffusion process can be considered as a sequence of \guillemotleft waiting times\guillemotright\ when the particle is in a quasi-stationary trapped state followed by \guillemotleft breakouts\guillemotright\ into ballistic motion until the next trapping and waiting period starts \citep[][]{shlesinger1987,klafter1990}. Such particle motions are typical, for instance, for {L\'evy flights} \citep[cf., e.g.,][]{shlesinger1993}. 

Working in $d$-dimensions, the probability of a particle to occupy a particular volume element  during a process, assumed to be caused by some unspecified (nonlinear) interaction between particles and plasma waves, is most conveniently formulated in wave number space $\mathbf{k}$ with probability spectrum 
\begin{equation}\label{eq-one}
p(\mathbf{k})\propto\exp(-ak^{\alpha}), 
\end{equation}
where $a$ is some positive constant, and $0<\alpha\in\textsf{R}$  a real number.  $\alpha\geq2$ reproduces the classical Gaussian probability spectra \citep{tsallis1995}. Non-Gaussian spectra have flatter tails implying $\alpha<2$, indicating superdiffusion. The connection of the above probability spectrum to real space distributions, in particular to $\kappa$ distributions, is non-trivial.

The diffusion process can be envisaged as consisting of a sequence of $n$ steps \citep[cf., e.g.,][]{treumann1997} bridging the time from $t=0$ to $t=t_n$ with the particle jumping from first waiting to $n$th waiting position, the expectation value of the latter becomes
\begin{equation}\label{eq-four}
\langle \mathbf{x}^2(n)\rangle =\int \mathbf{x}^2p(n\,|\,\mathbf{x})\mathrm{d}^d x, \qquad p(n)=\prod\limits_1^n p(i).
\end{equation}
The $n$th expectation value is proportional to the random mean square of the displacement $\mathbf{x}^2$ and a power of the elapsed time sequence. Working in Fourier (or momentum) space $\mathbf{k}$, multiplication of the probabilities yields 
\begin{equation}\label{eq-five}
p(n\,|\,\mathbf{k})=p^{n}(\mathbf{k})\propto \exp(-ank^\alpha) \sim \ p\,(\mathbf{k}')
\end{equation}
with $p\,(\mathbf{k}')$ the probability of the $n$th time step. Hence $\mathbf{k}'=\mathbf{k}n^{1/\alpha}$.
Any real space coordinate therefore scales as $x\to xn^{-1/\alpha}$. For the real-space probability this implies that
\begin{equation}\label{eq-seven}
p(n \,|\, \mathbf{x})\ \mathrm{d}^dx \ \longrightarrow\ p\big(\mathbf{x}/n^{1/\alpha}\big)\ \mathrm{d}^dx/n^{d/\alpha}
\end{equation}
yielding from Eq. (\ref{eq-four}) for the $n$th displacement expectation value 
\begin{equation}\label{eq-eight}
\big\langle \mathbf{x}^2(n)\big\rangle = n^{2/\alpha} \langle\mathbf{x}^2\rangle 
\end{equation}
with $\alpha<2$ not precisely known but to be determined below from numerical simulations. The mean-square displacement should be obtained from the second moment of the underlying real-space distribution function, for instance the $\kappa$ distribution, yielding
\begin{equation}\label{eq-ten}
\langle \mathbf{x}^2\rangle ={\textstyle\frac{1}{2}}d\kappa(\kappa+1)\ell^2
\end{equation}
an expression we will make tentative (not fully justified and for the present purposes marginal) use of only at the very end in 
application to reconnection.

\section{Diffusion coefficient} 
In using probability steps $n$, time has been discretized into pieces of free flight, waiting and some kind of interaction. In the average the interaction is covered by a fictitious anomalous collision frequency $\nu_a$.  Ordinary binary collision frequencies $\nu_c$ are very small, suggesting a scaling $\nu_a\gg \nu_c$ with the anomalous timescale $\nu_a^{-1}=\tau_a\ll\tau_c=\nu_c^{-1}$ much less than the collision timescale $\tau_c$. The diffusion process takes place in a time $t<\tau_c$. Replacing the time steps $n\to \nu_at$  the mean square $n$th displacement becomes 
\begin{equation}\label{eq-nine}
\langle\mathbf{x}^2(t)\rangle = \langle\mathbf{x}^2\rangle (\nu_at)^{2/\alpha}
\end{equation}
{With $\gamma=2/\alpha$ it defines the {anomalous} diffusion coefficient $D_a$
when multiplying by $\tau_a^{-1}=\nu_a$ 
\begin{equation}\label{da}
D_{a}(d, t)=\langle \mathbf{x}^2\rangle (\nu_at)^{2/\alpha}\nu_{a}\equiv D_{ca}(\nu_at)^{2/\alpha}
\end{equation}
as a function of time $t\nu_a$. Since $\nu_a\gg \nu_c$, it is much less than the classical diffusion coefficient which in this case would correspond to free flight. Under anomalous collisions the free flight is abruptly interrupted and reduced to non-stochastic  diffusion by the finite anomalous collision frequency $\nu_a$.}

 \begin{figure*}[t!]
\centerline{\includegraphics[width=0.9\textwidth,clip=]{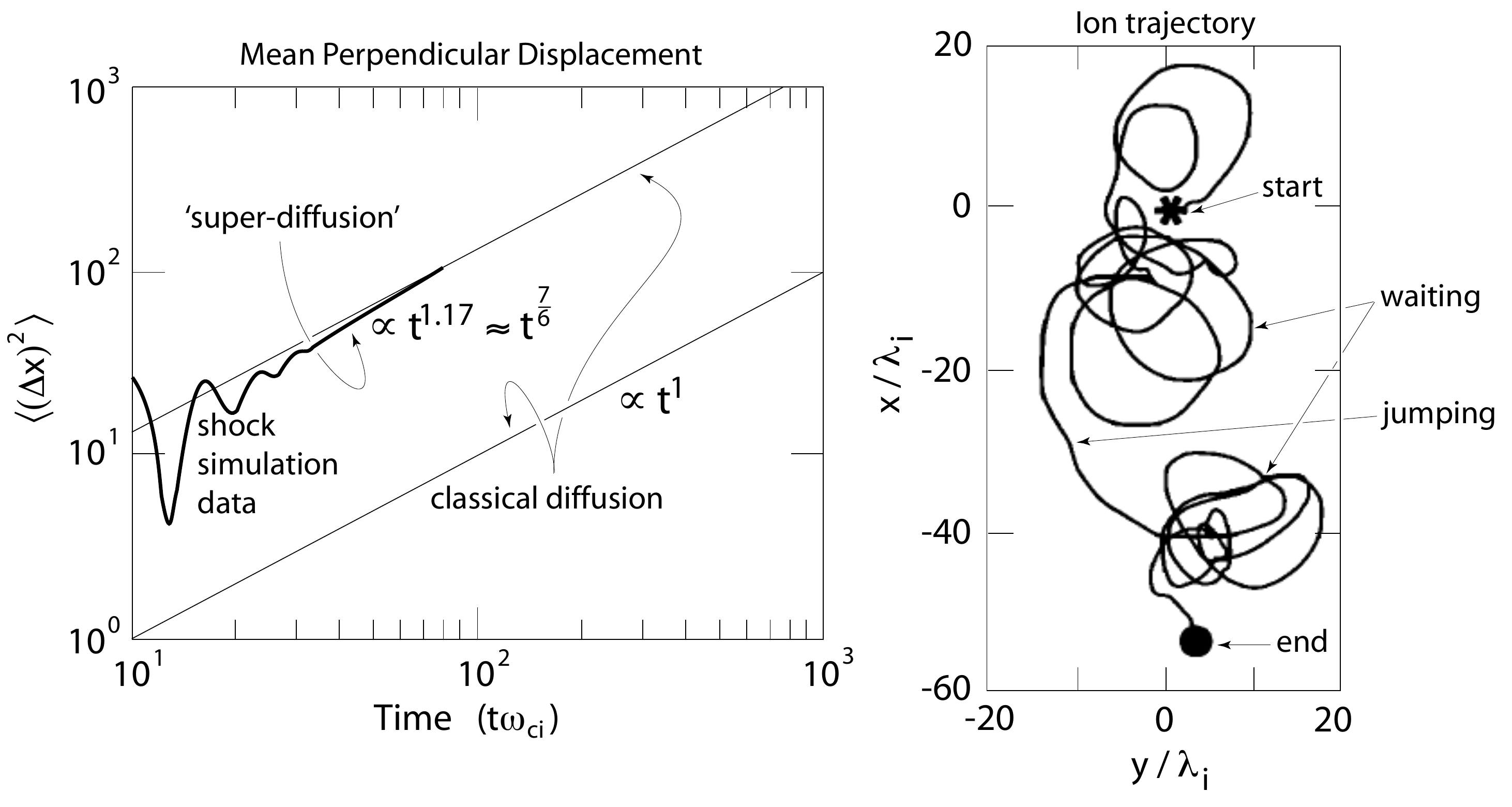} 
}
\caption[]
{Two-dimensional numerical simulation results of the mean downstream perpendicular displacement of ions near a quasi-perpendicular supercritical shock (shock normal angle $\theta=87^\circ$, Alfv\'enic Mach number $M_A=4$ as function of simulation time \citep[simulation data taken from][courtesy American Geophysical Union]{scholer2000}. Distances are measured in ion inertial lengths $\lambda_i=c/\omega_i$ with $\omega_i$ ion plasma frequency. \emph{Left}: The particle displacement performs an initial damped oscillation before settling into a continuous diffusive increase at time about $\omega_{ci}t\sim 40$ [in units of the ion gyro frequency $\omega_{ci}$]. The further time-evolution deviates apparently only slightly from the classical (linear) increase of the mean displacement, following a $\langle (\Delta x)^2\rangle\propto (\omega_{ci}t)^{1.17}$ power law. (Note that simulation-time limitations did not allow monitoring the long-time evolution of the ensemble-averaged square displacement, thus inhibiting determination of the final state of the diffusion process.) \emph{Right}: Late time trajectory of an arbitrary ion of the sample used. The orbit is projected into the plane perpendicular to the mean magnetic field which consists of a superposition of the ambient and wave magnetic fields. The ion orbit is neither an undisturbed gyro-oscillation nor a smooth stochastic trajectory. It consists of waiting (trapped gyrating) parts and parts when the ion suddenly jumps ahead a long distance cause by some brief but intense interaction between the particle and wave spectrum. This break out of gyration is typical for rare extreme events like those in L\'evy flights referred to in the present paper.}\label{fig-SPPS}
\end{figure*}

\section{Evolution}
Estimates of diffusion coefficients respectively $\gamma$ based on observations in space plasma are not only rare but unreliable. They suffer from the practical impossibility of any sufficiently precise determination of particle displacements as function of time and the subsequent transition to the asymptotic state. In addition they are mostly based on quasilinear theories of particular instabilities \citep{sagdeev1966,liewer1973,huba1977,davidson1978,sagdeev1979,huba1981,labelle1988,treumann1991,yoon2002,matthaeus2003,daughton2004,ricci2005,roytershteyn2012,izutsu2013} which do not properly account for any nonlinear interactions. 


We therefore refer to high-resolution particle-in-cell simulations \citep{scholer2000} performed in order to determine the  cross-magnetic field diffusion of ions near quasi-perpendicular shocks. The results are compiled in Figure \ref{fig-SPPS}. 

The right-hand side of the figure shows one macro-particle orbit arbitrarily selected out of the large number of particles used in the simulation to determine their instantaneous displacements from the origins of their trajectories in the simulation as function of simulation time measured in units of their identical (energy-independent) gyration frequency $\omega_{ci}=eB/m_i$ in the total magnetic field, which is the sum of the ambient and the self-consistently generated turbulent wave magnetic field. The particle shifts its position perpendicular to the magnetic field from its start point to the end point in the simulation. It is found in a slowly changing waiting position, performs jumps to new waiting positions, and ends up during a final jump. Such an orbit it neither adiabatic nor stochastic. 

The left part of the Figure shows the average displacement, ensemble averaged over the entire particle population, as function of simulation time. After performing an initial oscillation the average displacements settle into an about smooth continuously increasing curve of constant slope $\langle(\Delta x)^2\rangle \propto t^{1.17}$.

The slope of the final evolution of the average displacement is close to but by no means identical with classical diffusion which is shown by the slope of the two straight lines in the figure. Though the deviation in the slope is small, it is nevertheless substantial and statistically significant, indicating a superdiffusive process which deviates from classical diffusion. (We should note that, because of the large number of $\sim 6.3\times10^6$ macro-particles used in the simulation of which 525000 had high energies and contribute most to the mean displacement as well as for the high time resolution, the statistical error of the measurement is smaller than the width of the line in this figure!) 

{Adopting the probability spectrum based theory the experimentally determined slope of $2/\alpha \approx 1.17$ of the average displacement in Figure \ref{fig-SPPS} tells that in these simulations one had 
\begin{equation}
\alpha\approx 1.71 \qquad \mathrm{(experimental)}
\end{equation}
a value substantially far away from the Gaussian limit spectral slope $\alpha=2$ and being less than it, thus indicating quite strong superdiffusion. }

\section{Transition to collisional state}
Anomalous diffusion proceeds on a faster than classical time scale with time dependent diffusion coefficient which justifies the term superdiffusion. In spite of this, the coefficient $D_{ca}=\langle\mathbf{x}^2\rangle\nu_a$ in front of the time factor determining the absolute magnitude of the diffusion is generally small.  It does not compensate for the \emph{absolute} smallness of the diffusion coefficient. When, after a long time has elapsed the order of the classical collision time $\tau_c$, classical diffusion takes over scattering some particles to larger, some others back to smaller displacements and setting the collisionless process temporarily out of work. The average displacement of the violently scattered particles whose displacement line has been smeared out suddenly over a large spatial domain may now follow the classical linear temporary increase. 

\begin{figure*}[t!]
\centerline{\includegraphics[width=0.9\textwidth,clip=]{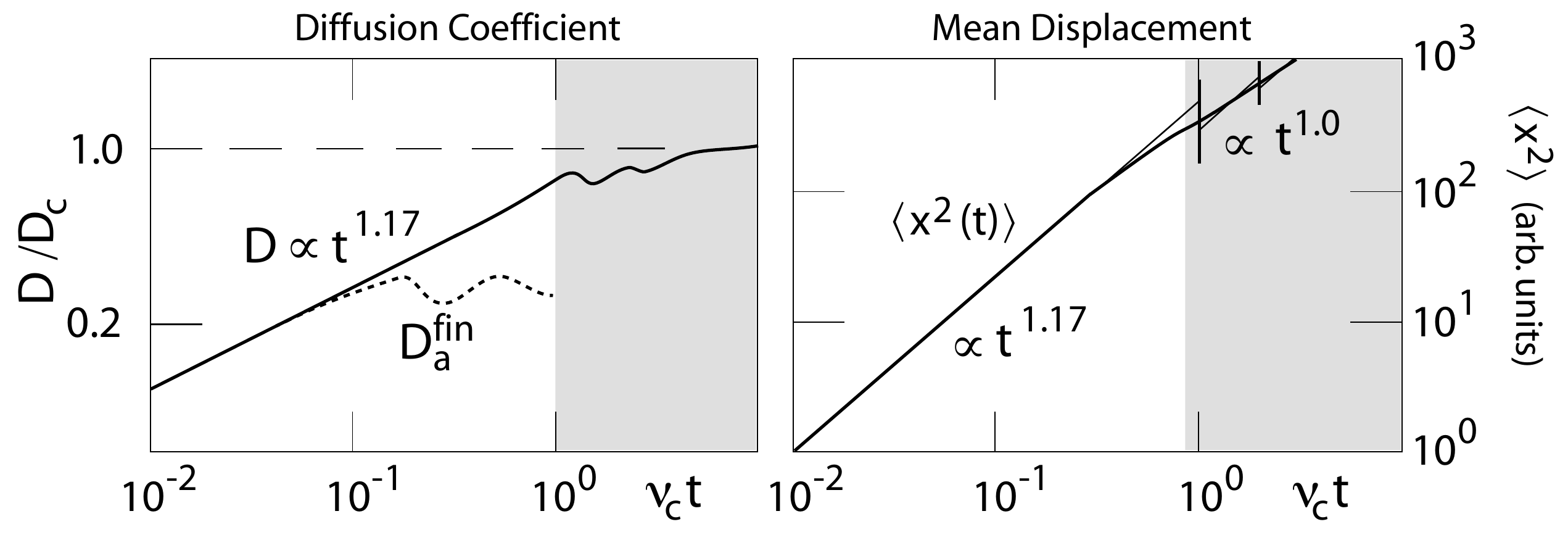} 
}
\caption[]
{Schematic hypothetical evolution of the diffusion coefficient for the case simulated in Figure 1 until the collisional classical diffusion state would have been reached. Time is measured here in classical collision times $\nu_c^{-1}$. \emph{Left}: The anomalous increase of the diffusion coefficient with time. The growth of the diffusion coefficient gradually comes to rest after the classical collision time has elapsed. \emph{Dotted}: A time dependent nonlinear stationary state never approaching classical diffusion. \emph{Right}: Time evolution of the average particle displacement increasing like shown in Figure 1. When approaching the classical collision time, scattering of particles to both larger and smaller displacements widen the displacement range, leading to a reduced increase until the second collision time. Similarly after the second, third, and the following collision times. Finally, the increase of the displacement settles into linear in time, implying classical or stationary diffusion. }\label{fig-andiff}
\end{figure*}

One single elapsed binary collision time may not suffice to stop the nonlinear collisionless interaction process. The widely scattered particle population may still have sufficient freedom to organize again into a softened collisionless diffusion which lasts until the next binary collision time has passed. During this second collisionless period the slope should be flatter than the initial collisionless slope, and after statistically sufficiently many periods of elapsed classical collision times no collisionless mechanisms revives anymore. Diffusion has by then become completely classical. These  sequences are schematically shown in Figure \ref{fig-andiff}. 

\section{Discussion} 
Waiting statistics offers an approach to anomalous diffusion in various regions of space plasmas where classical (and neo-classical) diffusion processes are inappropriate, violently failing to explain the transport of plasma and magnetic field. Application to numerical simulations near collisionless shocks determined the value of $\alpha\approx 1.71$ which turns out to be close to but sufficiently far below its classical (Gaussian) limit $\alpha=2$ for identifying superdiffusion. Superdiffusion coefficients obtained are small but increase with time. 

The present theory is based on constant $\alpha$ for the entire diffusion process. This might be unrealistic. Real powers $\alpha\, [W_w(t)]$ will turn out functionals of  the time-dependent turbulent wave levels $W_w(t)$ which are generated self-consistently in the underlying turbulent collisionless wave-particle interaction \citep[for a derivation of the phase-space distribution in particular wave-particle interactions cf., e.g.,][yielding time-asymptotic values of the phase-space power-law index $\kappa$ depending on wave power $W_w$]{hasegawa1985,yoon2012}. 

It may be expected that, with increasing wave level $W_w(t)$, a new collisionless equilibrium will be reached where the diffusion process, in finite time $t\sim\tau_f$, approaches another new and approximately constant diffusivity 
\begin{equation}
\lim\limits_{\textstyle t\to\tau_f}D_a(t)\ {\textstyle\longrightarrow}\ D_a^\mathit{fin}(t\gtrsim\tau_{f})<D_c 
\end{equation}
for $\nu_a^{-1}(t=0) \lesssim \tau_\mathit{f} \ll \nu_c^{-1}$, with $W_w(t\gtrsim\tau_f),\ \alpha\,[W_w(t\gtrsim\tau_f)]$ both either constant or oscillating around their time-averaged mean values $\big\langle W_w(t\gtrsim\tau_f)\big\rangle,\big\langle\alpha(t\gtrsim\tau_f)\big\rangle$, and the final average diffusion coefficient $\big\langle D_a^\mathit{fin}(\tau_f\lesssim t\lesssim \nu_c^{-1})\big\rangle$ remaining constant. {Under such circumstances the diffusion coefficient in Figure 2 never approaches the classical limit but settles instead on its much lower anomalous collisionless level $\big\langle D_a^\mathit{fin}\big\rangle$. The related processes lie outside the present investigation. We may, however, estimate a lower bound on the average final diffusion coefficient $\left\langle D_a^\mathit{fin}\right\rangle$ assuming $\tau_f\approx\nu_a^{-1}$, which yields 
\begin{equation}
D_\mathit{ca}\lesssim  \left\langle D_a^\mathit{fin}\right\rangle
\end{equation}}

In the following we list a few practical consequences of our theory which focus on one of the most interesting problems in collisionless plasma physics, the mechanism of collisionless reconnection of magnetic fields.

\subsection{Resistive scale and relation to reconnection}
We may use these arguments to briefly infer about the resistive scale $L_\nu$, a quantity frequently referred to in discussions of diffusion in presence of current flow. It plays a role in the diffusive evolution of the magnetic field which from the induction equation is given in its simplest form
\begin{equation}
\frac{\partial\mathbf{B}}{\partial t}=\nabla\times\mathbf{V\times B}+D_m\nabla^2\mathbf{B}, \qquad D_m=\frac{\eta}{\mu_0}=\lambda_e^2\nu
\end{equation}
The resistive scale is defined as $L_\nu^2\sim D_mt=\lambda_e^2\nu t$ being determined through resistivity $\eta=\nu/\epsilon_0\omega_e^2$ and electron inertial length $\lambda_e=c/\omega_e$, with plasma frequency $\omega_e$. It tells, at what scale resistive diffusion starts affecting the plasma dynamics.

{It is interesting to know how the resistive scale evolves with time in a nonlinearly active though collisionless medium. Using the expression for the product $\nu_at$ to replace $\nu t$ gives
\begin{equation}
\frac{L_{\nu_a}}{\lambda_e}\sim \left\{\frac{D_{a}(t)}{D_{ca}}\right\}^{1/2} \sim \left(\nu_a t\right)^{1/\alpha}
\end{equation}
for the resistive scale in units of $\lambda_e$, expressed through the (time dependent) diffusion coefficient $D_{a}$. This indicates that the resistive scale increases with time from a value $L_{\nu_a}<\lambda_e$ until $D_{a}\sim\left\langle D_a^\mathit{fin}\right\rangle$ when $L_{\nu_a}^\mathit{fin}\sim\lambda_e$ approaches the inertial scale.}

{Small (anomalous) resistive scales imply fast magnetic diffusion as observed in collisionless systems like in reconnection. Since in collisionless plasma there is no resistive diffusion, one concludes that any process causing diffusion will readily reduce the resistive scale to values below the electron inertial scale causing comparably fast dissipation of magnetic fields and favoring reconnection.} 

The remaining problem consists in finding an appropriate expression for the \emph{equivalent} anomalous \guillemotleft collision frequency\guillemotright\ $\nu_a$ under collisionless conditions. Observations \citep{labelle1988,treumann1990,bale2002} do not indicate any presence of sufficiently high wave amplitudes in collisionless reconnection required \citep{sagdeev1966,sagdeev1979} for the quasilinear generation of anomalous resistances. Numerical particle-in-cell simulations \citep[cf.,][for a recent review]{treumann2013} confirmed instead that in all cases the main driver of fast collisionless reconnection is the electron \guillemotleft pseudo-viscosity\guillemotright\ implied by the presence of non-diagonal terms \citep{hesse1998,hesse1999} in the thermally anisotropic electron pressure tensor $\textsf{P}_e$ measured in the stationary frame of the reconnecting current layer and accounting for any subtle finite gyro-radius effects in the dynamics of electrons in the inhomogeneous magnetic field of the electron diffusion region where electrons perform bouncing Speiser orbits. 

\subsection{Gyroviscosity}
An expression for the  anomalous collision frequency $\nu_a$ that is \emph{equivalent} to electron pseudo-viscosity is found referring to  the volume viscosity $\mu_V$ (or kinematic viscosity $\mu_\mathit{kin}=\mu_V/mN$, with $N$ the density) and the molecular collision frequency $\nu_m$ \citep[][]{huang1987} 
\begin{equation}
\mu_V=NT/\nu_m \qquad\mathrm{or}\qquad \mu_\mathit{kin}=T/m\nu_m
\end{equation}
Formally, this allows for the determination of  $\nu_a$ when identifying $\mu_V$ with the electron volume \guillemotleft pseudo-viscosity\guillemotright \ $\mu_e$ (or kinematic pseudo-viscosity $\mu_{e,kin}=\mu_e/Nm_e$) resulting from the non-diagonal electron pressure tensor elements, a quantity which can be determined either from observation or from numerical particle-in-cell simulations. This yields
\begin{equation}
\nu_a\approx NT_e/\mu_e=T_e/m_e\mu_{e,\mathit{kin}}
\end{equation}
with $N$ the plasma density and $T_e$ the relevant electron temperature for the \emph{pressure-tensor induced equivalent} anomalous collision frequency.  \citet{macmahon1965} derived an MHD form of the full pressure tensor including finite ion-gyroradius contributions {in the limit of very strong magnetic fields}, barely applicable to the weak magnetic field reconnection site. A simplified version of his expressions neglecting heat fluxes was given by \citet{stasi1987} based on the implicit assumption that in strong magnetic fields the mean free path is replaced by the ion-gyroradius. In view of reconnection, this form has been used by \citet{hau1991} in application to rotational discontinuities \citep[for the role of viscosities in viscous fluids cf.][]{landau1959}. 

In this form, rewritten for the relevant electron dynamics, one has $\mu_e\simeq T_e/m_e\omega_{ce}$, which identifies $\nu_a=\nu_{gv}\sim\omega_{ce}$ as an {electron gyro-viscous} MHD collision frequency of the order of the electron cyclotron frequency $\omega_{ce}=eB/m_e$ -- indeed much larger than any Coulomb collision frequency. It suggests that gyro-viscous superdiffusion means Bohm diffusion. 

\subsection{Estimates of transport quantities}
Instead, use can be made of available numerical simulations  \citep{pritchett2005} which quantitatively determined the contribution of the electron-pressure tensor-induced pseudo-viscosity to the dissipative generation of the parallel electric field in guide-field reconnection \citep[cf.,][for a critical discussion]{treumann2013}. \citet{pritchett2005} obtained for the maximum non-diagonal pressure-generated field $E_{\|,P}$ in the inner part of the reconnection site (or electron exhaust region)
\begin{equation}
E_{\|,P}\ =\ (eN)^{-1}\left|\nabla\cdot\textsf{P}_e\right|\ \ \lesssim\ \ 0.4\ V_{A}B_0
\end{equation}
where $N,B_0,V_{A}$ are the respective density, magnetic field outside the current layer, and Alfv\'en velocity based on $B_0$.  The width of the current layer was $L_s\sim 2\lambda_i=2\sqrt{M_s}\lambda_{es}$, with simulation mass ratio $M_s=m_i/m_{es}=64$. On using index $s$ for simulation quantities, real electron masses become $m_{e}=r\, m_{es}$, with $r=64/1840$. With current $\mathbf{J}$, we may put
\begin{equation}
E_{\|,P}\ =\ \eta_{as} |\mathbf{J}|\ \sim\ \frac{\eta_{as}}{\mu_0} \frac{B_0}{L_s}\ =\ \frac{\lambda_{es}\nu_{as} B_0}{2\sqrt{M_s}}
\end{equation}
Thus, the anomalous collision frequency corresponding to the pressure induced pseudo-viscosity in the simulation of the reconnection process was of the order of
\begin{equation}
\nu_{as}\ \lesssim\ \ 0.8\ \ \sqrt{M_s}\ (V_{A}/c)\ \omega_{es}\ =\ \, 0.8\  \omega_{ce,s}
\end{equation}
with the second form of the right-hand side resulting when accounting for the identity $(V_A/c)\sqrt{M}=\omega_{ce}/\omega_e$.  In terms of real electron masses the last expression becomes
\begin{equation}
\nu_{a}\ = \nu_{as} r \ \lesssim\ 0.03\ \omega_{ce}
\end{equation}
This value is more than one order of magnitude smaller than the one of $\nu_{gv}$ obtained above from gyro-viscous  MHD theory, rewritten for electrons. Still, its value is uncertain for the unknown dependence on mass ratio of the reconnection electric field $E_{\|,P}$ in the simulations. Assuming that this dependence is moderate, the agreement is surprisingly reasonable. For the wanted pseudo-viscosity this gives 
\begin{equation}
\mu_{e,kin}\approx T_e/m_{es}\nu_{as}=T_e/m_e\nu_a \gtrsim 1.25\ T_e/m_e \omega_{ce}
\end{equation}
with the factor $r$ in the denominator canceling, a form similar to gyro-viscosity for both simulation and real plasma applications. 

Adopting the above numerical estimate of $\nu_a$,  the anomalous diffusion coefficient 
\begin{equation}\label{supd}
D_{a}(t) = 1.65\times10^{-2} D_{ca}\ (\omega_\mathit{ce}t)^{1.17}
\end{equation}
increases slowly with time measured in electron cyclotron periods. 

\subsection{Digression on $\kappa$}
With the last formula we have, in principle, achieved our goal. 

However, someone might want to know the explicit form of the diffusion coefficient. For this one needs to determine the coefficient $D_{ca}$, which requires knowledge of $\langle\mathbf x^2\rangle$ in the electron exhaust. Since, from the simulations, no information is available on displacements, one has to refer to model assumptions for the distribution function $p(\mathbf{x})$.  

Among the limited number of such functions available one may adopt the $\kappa$ distribution Eq. (\ref{eq-two}), even though it is rather improbable that in the tiny reconnection region and for the restricted reconnection time any stationary $\kappa$ distributions will have sufficient time to evolve. 

Nevertheless, in the absence of any better choice, one may tentatively evoke the relation $\alpha/2=\kappa(\kappa+d/2)^{-1}$ between $\alpha$ and $\kappa$, as proposed from non-extensive statistical mechanics \citep{tsallis1995,prato1999,bologna2000,liva2013} to hold in the superdiffusion range $\alpha<2$, and apply it as well to our particular reconnection problem. 

Then, on using the measured value of $\alpha$, we have $\kappa\approx 5.9$ for $d=2$. This gives the two-dimensional $\kappa$-superdiffusion coefficient from Eqs. (\ref{supd}), (\ref{eq-ten}), and (\ref{da}), with squared correlation length $\ell^2=2T_e/m_e\nu_a$, as
\begin{equation}\label{dkap}
D_{a\kappa}(t)\approx 11\ D_B\left(\omega_{ce}t\right)^{1.17}
\end{equation}
where $D_B\approx T_e/m_e\omega_{ce}$ is of the order of the Bohm diffusion coefficient. This value of ten times (!) Bohm diffusion is excessively large, implying the presence of extraordinarily strong anomalous diffusion at the reconnection site though being not in unacceptable disagreement with exceptionally fast spontaneous reconnection. For a Gaussian probability distribution one had $\langle x^2\rangle= \ell^2 d/2$ and thus $D_a(t)\approx D_B\left(\omega_{ce}t\right)^{1.17}$. 

It should, however, be kept in mind that the derivation of the $\kappa$-diffusion coefficient Eq. (\ref{dkap}) is based on the arbitrary assumption that the unknown distribution of displacements in the narrow electron exhaust would indeed be of the family of $\kappa$ distributions. While the determination of the anomalous collision frequency from the simulations used is very well justified, there is no observational, nor any theoretical, justification for this \emph{ad hoc} assertion, however. 

\subsection{Lower limit on $\nu_a$ in reconnection}
{The above numerical simulation based estimates can be directly applied to observations of reconnection in the magnetotail current sheet in order to infer about the anomalous collision frequency generated in reconnection. From an applicational geophysical point of view this is most interesting. Observed magnetic fields across the tail plasma sheet vary between $1\ \mathrm{nT} < B_0 < 10\ \mathrm{nT}$. With these values one obtains the following range for the anomalous collision frequencies during reconnection in the plasma sheet:
\begin{equation}
4.9\ \mathrm{Hz}\ < \ \nu_{a\ }\  < \ 50\ \mathrm{Hz},\qquad\ \ \, \omega_{\,lh\ }\ \,\approx \ \ 4.1\ \ \mathrm{Hz}
\end{equation}
These reasonably high values follow directly from analysis of the simulations, compared to the lower-hybrid frequency $\omega_{lh}$ given on the right  for the lower value $B=1$\,nT only. This estimated anomalous collision frequency at the magnetotail reconnection site is the result of non-stochastic processes in the electron exhaust diffusion region which generate the out-of diagonal pseudo-viscous terms in the electron pressure tensor. It is responsible for the necessary  superdiffusion at the reconnection site which is required in the collisionless reconnection process. } 

The closeness of the lower-hybrid frequency $\omega_\mathit{lh}$ to the range of anomalous collision frequencies indicates the collisionless electric coupling between electrons and ions in any reconnection process. 

In addition, it provides an important {lower limit}
\begin{equation}
\omega_{lh}\,\lesssim\,\min_\mathit{rec}\,(\nu_a)
\end{equation}
on $\nu_a$ in collisionless reconnection, thereby  \emph{a posteriori} justifying the frequently found surprising closeness \citep[e.g.,][and others]{huba1977,labelle1988,treumann1991,yoon2002}  to the lower-hybrid frequency of the rough estimates of anomalous collision frequencies  from the analysis of spacecraft observations of reconnection which are necessary to explain the time scale of the observed dissipation of energy. 

Considered in this spirit, collisionless reconnection is understood as an \emph{equivalent anomalous local super}-diffusion process in collisionless plasma. From a general physical point of view, this interpretation ultimately re-unifies the initially considered mutually excluding collisionless reconnection and diffusion theories in satisfactory concordance with fundamental electrodynamics.

\begin{acknowledgements}
This research was part of a Visiting Scientist Program at ISSI, Bern. Hospitality of the librarians Andrea Fischer and Irmela Schweizer, and the technical administrator Saliba F. Saliba,  is acknowledged. RT thanks the referees for clarifying comments and suggestions of related literature.
\end{acknowledgements}











\appendix
{\section{\\ \\ \hspace*{-7mm} Anomalous spectra and $\kappa$ distributions}}
Based on semi-quantitative asymptotic arguments it has been argued \citep{tsallis1995,prato1999,bologna2000,gell-mann2004,liva2013} that $q$- and $\kappa$-distributions both belonged to the class of anomalous L\'evy-like $\alpha$-probability spectra  Eq. (\ref{eq-one}) rendering valid a relation between $\alpha$ and $\kappa$ of the kind used in subsection 6.4 in the present paper. 

Below we show by rigorous calculation in two different ways  that these arguments seem doubtful. Apparently $q$ and $\kappa$ probability distributions do not belong to this kind of L\'evy-like $\alpha$-spectra. Their spectral form is substantially more complicated.\footnote{See also endnote No 34 in \citet{bologna2000} where it is explicitly noted that problems remain with the relation between L\'evy spectra and $q$ distributions, i.e. an unambiguous $\alpha[q]$ relation.}   
\\

{\subsection*{Probability distribution from $k^\alpha$ spectrum}  
Retransforming the probability spectrum Eq. (\ref{eq-one}) into real space requires solving the inverse Fourier integral
\begin{equation}
p(\alpha\,|\,\mathbf{x})\propto \frac{1}{(2\pi)^3}\int \mathrm{e}^{-ak^\alpha -i\mathbf{k\cdot x}}\mathrm{d}^dk
\end{equation}
in $d$ dimensions and properly normalized. Here $\ell k\to k, \ x/\ell\to x, \ a/\ell^\alpha\to a$. No general solution is known for this integral except in the case $\alpha=2$. Its solution for arbitrary real $\alpha\in\textsf{R}$ can be attempted  applying the method of steepest descent. Aligning $\mathbf{k}$ and $\mathbf{x}$, one has $\mathrm{d}^3{k}=k^{d-1}\mathrm{d}k$. Unfortunately, the turning point equation
\begin{equation}
k_\mathit{tp}^\alpha+ixk_\mathit{tp}/a\alpha +(d-1)/a\alpha = 0
\end{equation}
cannot be solved for arbitrary $\alpha$. The only two treatable cases are $\alpha=2$ for all dimensions $d$, and the one-dimensional case $d=1$.  Only the latter is of interest. \citep[It is well-known, cf., e.g.,][for a recollection, that the former trivially reproduces the Gaussian distribution.]{tsallis1995} One thus has $k_{\mathit{tp},d=1}=(-ix/a\alpha)^{1/(\alpha-1)}=(x/a\alpha)^{1/(\alpha-1)}\exp\left[3\pi i/2(\alpha-1)\right]$, and with $d=1$ for the above Fourier integral
\begin{equation}
p(\alpha\,|\,\mathbf{x})\propto\frac{\mathrm{e}^{-ak_\mathit{tp}^\alpha-ik_\mathit{tp}x}}{2\pi}\int \mathrm{d}k\ \mathrm{e}^{-\frac{1}{2}a\alpha k_\mathit{tp}^{\alpha-2}(k-k_\mathit{tp})^2}
\end{equation}
The condition that $\Re(k_\mathit{tp}^{\alpha-2})>0$ yields the trivial requirement $\alpha<5/2$ valid for all interesting cases including the Gaussian.} 

{Another condition is obtained from the requirement that $p(\alpha\,|\,\mathbf{x})$ must be a real probability distribution. Setting $\Im(k_\mathit{tp})=0$ one concludes that $\alpha_{d=1}=1+3/2n\ >\ 1$, with $n=1,2,\dots$. Hence, $\alpha_{d=1}=\frac{5}{2},\frac{7}{4},\frac{3}{2},\frac{11}{8},\dots$ can assume discrete values only which, for large $n$, converge to 1.} 

{For arbitrary $d$ and $\alpha=2-\alpha'\approx 2$, i.e. $\alpha'\ll 1$, the two approximate solutions for the turning point become both purely imaginary $k_\mathit{tp}^{(1,2)}=-i[2(d-1)/x\ ;\  x(1-\alpha'/2)/2a -4(d-1)/x]$. The reality condition for the real space probability requires treating the complex turning point integral. Even close to the Gaussian limit $\alpha=2$ calculation of the real space probability distribution from the hypothetical non-Gaussian spectrum Eq. (\ref{eq-one}) is nontrivial, causing serious doubts in the assumed generality of the $k^\alpha$ model spectrum. No further useful information is obtained. }

{The $x$-dependence of the solution of the $d=1$ integral is contained in the factor $\left(a\alpha/x\right)^{(\alpha-2)/2(\alpha-1)}$ which may be interpreted as the large-$x$ limit of the $\kappa$ distribution for $d=1$. It can, by comparison, be reconciled only for values $0<\alpha_{d=1}=(1-\kappa)/(\frac{3}{4}-\kappa)<\frac{5}{2}$, yielding  $0<\kappa < \frac{3}{4}$, a narrow and extreme range only at this dimensionality. The lower limit on $\kappa$ gives $\frac{4}{3}\lesssim\alpha_{d=1}$. Imposing the Gaussian limit $\alpha=2$ yields finally $0<\kappa\leq\frac{1}{2}$. This is rather different from the Gaussian limit $\kappa\to\infty$ of the $\kappa$-distribution, in principle suggesting that the $\kappa$-distribution is \emph{not a good descriptive model} of hypothetical probability power spectra of the type of Eq. (\ref{eq-one}). This discussion holds for $d=1$. Little is known about the behavior at larger dimensions.}


{\subsection*{Spectrum of $\kappa$ distribution}}
{Solving the Fourier integral of the $\kappa$-distribution in $d$ dimensions requires requires calculating the integral
\begin{equation}
p_\kappa(\mathbf{k})\propto A_\kappa D^{(d-1)}\frac{\sqrt{\kappa}}{(-i\bar{k})^{d-1}}\int\limits_{-\infty}^\infty p(\kappa| \mathbf{\bar x})\mathrm{e}^{-i \bar{k}\bar{x}}\mathrm{d}\bar{x}
\end{equation}
where $\bar{k}=\sqrt{\kappa}\ell k,\ \bar{x}=x/\sqrt{\kappa}\ell$, and $D^{(d-1)}=\partial^{d-1}/\partial\bar{k}^{d-1}$. The solution is
\begin{eqnarray}\label{eq-powerkappa}
p_\kappa(\mathbf{k})&\propto& A_\kappa D^{(d-1)} 
 \frac{\sqrt{\kappa}\ \bar{k}^{\kappa+1-d/2}}{2^{\kappa+1+d/2}(-i)^{d-1}}  \times\\
 &\times&\frac{\sqrt{\bar{k}/\pi}}{\Gamma(\kappa+1+d/2)}K_{\frac{1}{2}(\kappa+1+d/2)}\left(\textstyle{\frac{1}{2}}\bar{k}\right)\nonumber
\end{eqnarray}
It is the large argument form of the Bessel function $\sqrt{\bar{k}/\pi}\ K_\nu(\bar{k}/2) \propto\ \exp(-\bar{k}/2)$ which contributes the wanted exponential factor. However, one finds that $\alpha\approx1$ is independent of $\kappa$ in this case, at least for the long scales. \citep[A similar conclusion was presented already in][]{treumann1997}. This behavior is not changed by the dimensional derivatives because of the reproductive property of the exponential. For $d=1$, in particular, the derivative disappears and the spectrum is completely described by the Bessel function. Physically it indicates that in the long range the spectrum corresponding to the $\kappa$ distribution is indeed, as expected, far away from Gaussian behavior. It, however, has no similarity whatsoever to the model probability spectrum in Eq. (\ref{eq-one}). The $\kappa$ distribution describes strong correlations between the particles with, if at all, weak dependence of $\alpha$ on $\kappa$ (at the best to higher order).}

{In order to get a feeling of such a weak dependence one may manipulate the exponential contributed by the Bessel function and its frontal $\bar{k}$ dependent factor in the above expression for large $\bar{k}$ to become
\begin{eqnarray}
\bar{k}^\zeta\exp-(\bar{k}/2)&=&\exp\left[-{\textstyle\frac{1}{2}}\bar{k}\left(1-\ln\,\bar{k}^{2\zeta/\bar{k}}\right)\right] \nonumber\\ 
&\approx& \exp\left[-{\textstyle\frac{1}{2}}\bar{k}\exp\left(-\ln\,\bar{k}^{2\zeta/\bar{k}}\right)\right] \\
&\approx& \exp\left[-{\textstyle\frac{1}{2}}\bar{k}^{(1-2\zeta/\bar{k})}\right]Ê\nonumber
\end{eqnarray}
Again, this exponential reproduces when carrying out the differentiations. Hence one finds approximately for the functional
\begin{equation}\label{eq-functional}
\alpha[\kappa] \approx 1-2\zeta/\bar{k}, \qquad \zeta= \kappa+1-d/2
\end{equation}
depending on the dimensionality $d$ and the spectral scale $\bar{k}$. For $d=1$ it is independent of the differential operation. It is obvious that $\alpha=2$, the Gaussian limit, is not contained in this condition which shows that $\alpha$ is small and for large $\bar{k}$ is of order $\alpha\sim O(1)$, as already inferred above. Moreover, for large $\kappa$ one has from this expression $\alpha\approx 1-\sqrt{\kappa}/k$ explicitly excluding the limit $\kappa\to\infty$ unless uniformly also $k\to\infty$.}

{The last condition is satisfied for $\bar{k}>2\zeta$ only providing another condition holds,
\begin{equation}
0<\kappa< \frac{k^2}{16\ell^2}\left[1\pm\sqrt{1-\frac{4\ell}{k}\left(1-\frac{d}{2}\right)}\right]^2 \approx \left(\frac{k}{2\ell}\right)^2
\end{equation}
which, for instance, yields $0<\kappa\sim 1$ for the modest requirement $k/\ell\sim 2$, thus identifying a sufficiently large range of validity of $\kappa$  for all three spatial dimensions in treating the average displacement problem. No independent relation between $\alpha$ and $\kappa$ has yet been obtained.} 

{The only way of including the Gaussian limit $\alpha=2$ is via application of renormalization group methods to the functional Eq. (\ref{eq-functional}). This procedure yields
\begin{equation}
\alpha[\kappa]\approx 2 +\exp[-(1+2\zeta/\bar{k})]
\end{equation}
which for fixed $k$ and large $\kappa$ exhibits a dependence $\alpha\sim\exp(-\sqrt{\kappa}/k)$. This form indeed includes the limit $\lim\limits_{\kappa\to\infty}\alpha=2$. However, it is not of any interest as this limit is approached from above, yielding $\alpha\gtrsim 2$ for all $\kappa$.}

{One may notice that the exponential power spectrum is not only approximate here, it has not even been imposed nor used in the calculation of the power spectrum belonging to the $\kappa$ distribution. Hence, if the $\kappa$ distribution is believed to describe power spectra of the type of Eq. (\ref{eq-one}) invluding the underlying physical processes, then this can only be taken as very approximate.}

Generally speaking, though the $\kappa$-distribution contains the Gaussian probability for $\kappa\to\infty$, the reconstruction of a simple Gaussian probability spectrum from the $\kappa$ distribution is not successful. \newpage This is finally seen when straightforwardly taking the large order expansion of the Bessel function for  $\kappa\gg 1$. Then the above spectrum asymptotically behaves as 
\begin{equation}
p_\kappa(\mathbf{k}) \sim D^{(d-1)}\frac{\exp{-\bar\kappa\eta}}{\sqrt{\bar\kappa}(1+\bar{k}^2/4\bar\kappa^2)^{1/4}}
\end{equation}
with $\bar\kappa=\frac{1}{2}(\kappa+1+d/2)$, and $\eta=\sqrt{1+\bar{k}^2/4\bar\kappa^2}+\ln\left[(\bar{k}/2\bar\kappa)\left(1+\sqrt{1+\bar{k}^2/4\bar\kappa^2}\right)\right]$. 
It is not obvious how this spectrum matches an $\exp(-ak^2)$ function for $\kappa\to\infty$. 

Taking $\kappa \gg k$ yields a dependence $\sim \exp(-\bar\kappa-\bar{k}^2/8\bar\kappa)\to 0$ for $\kappa\to\infty$. On the other hand, in the limit $k\gg\kappa$ for all fixed $\kappa$, one has $\eta\sim \bar{k}/2\bar\kappa$ and thus a dependence $\sim \exp -\bar{k}/2$. 

This $\alpha=1$ spectrum had been elucidated already above and also in \citet{treumann1997}. It represents the asymptotic spectral behavior of $\kappa$ distributions, indicating that asymptotically these are practically independent of $\kappa$ and thus, at the best, are extreme versions $\alpha=1$ of spectra of the form of Eq. (\ref{eq-one}), indicating the presence of extraordinarily strong correlations.

\end{document}